# The stability of Twitter metrics: A study on unavailable Twitter mentions of scientific publications

Zhichao Fang[1*], Jonathan Dudek[1], Rodrigo Costas[1,2]

* Corresponding author

Zhichao Fang (ORCID: 0000-0002-3802-2227)

[1] Centre for Science and Technology Studies (CWTS), Leiden University, Leiden, The Netherlands.

E-mail: z.fang@cwts.leidenuniv.nl

Jonathan Dudek (ORCID: 0000-0003-2031-4616)

[1] Centre for Science and Technology Studies (CWTS), Leiden University, Leiden, The Netherlands.

E-mail: j.dudek@cwts.leidenuniv.nl

Rodrigo Costas (ORCID: 0000-0002-7465-6462)

[1] Centre for Science and Technology Studies (CWTS), Leiden University, Leiden, The Netherlands.

[2] DST-NRF Centre of Excellence in Scientometrics and Science, Technology and Innovation Policy, Stellenbosch University, Stellenbosch, South Africa.

E-mail: rcostas@cwts.leidenuniv.nl

**Abstract**

This paper investigates the stability of Twitter counts of scientific publications over time. For this, we conducted an analysis of the availability statuses of over 2.6 million Twitter mentions received by the 1,154 most tweeted scientific publications recorded by Altmetric.com up to October 2017. Results show that of the Twitter mentions for these highly tweeted publications, about 14.3% have become unavailable by April 2019. Deletion of tweets by users is the main reason for unavailability, followed by suspension and protection of Twitter user accounts. This study proposes two measures for describing the Twitter dissemination structures of publications: *Degree of Originality* (i.e., the proportion of original tweets received by a paper) and *Degree of Concentration* (i.e., the degree to which retweets concentrate on a single original tweet). Twitter metrics of publications with relatively low *Degree of Originality* and relatively high *Degree of Concentration* are observed to be at greater risk of becoming unstable due to the potential disappearance of their Twitter mentions. In light of these results, we emphasize the importance of paying attention to the potential risk of unstable Twitter counts, and the significance of identifying the different Twitter dissemination structures when studying the Twitter metrics of scientific publications.

## Introduction

Twitter has become one of the most important dissemination tools of scientific information and scholarly communication, used not only by the scientific community, but also by the public in general (Van Noorden, 2014; Kahle, Sharon, & Baram-Tsabari, 2016). Twitter is also one of the most predominant altmetric data sources for scientific publications (Robinson-Garcia et al., 2014; Haustein, 2019). Several studies have discussed aspects of data coverage, density, and intensity (Thelwall et al., 2013a; Haustein, Costas, & Larivière, 2015), or the accumulation velocity of tweets to publications (Fang & Costas, 2018). It is assumed that Twitter mentions, as well as other types of social media metric data, are more likely to measure a broader impact of research that differs from the academic impact reflected by citations (Bornmann, 2015; Robinson-Garcia, van Leeuwen, & Ràfols, 2018). Therefore, Twitter metrics are usually calculated with the motivation of further application in research assessment and science policy (Wilsdon et al., 2015; Haustein, 2019). In this context, the stability of metrics can be seen as a key component of data quality, being of great significance for a reasonable and sustainable measurement of the reception and discussion of research outputs on Twitter.

*Development of Twitter metrics*

The characteristics of altmetric data, such as broadness, speed, openness, and transparency (Wouters & Costas, 2012), have raised expectations towards the development of alternative indicators that can measure research impact in an early stage following publication (Priem & Hemminger, 2010). As a result, numerous studies have analyzed the correlation between various altmetric indicators and citation-based indicators, testing whether the former might be applied for predicting highly cited papers – which otherwise is impaired owing to the citation delay (Priem, Piwowar, & Hemminger, 2012; Zahedi, Costas, & Wouters, 2014; Waltman & Costas, 2014; Costas, Zahedi, & Wouters, 2015; Zahedi, Costas, & Wouters, 2017). As a source that contributed a considerable share of data about online activities associated with scholarly outputs only second to Mendeley (Sugimoto et al., 2017), Twitter has been widely discussed in previous research. Herein, the impact of scientific publications on Twitter was usually measured by counting the total number of mentions they have received or the total number of Twitter users who have mentioned them in their tweets. These two counting methods of Twitter metrics are commonly employed by altmetric data aggregators.

In spite of this strong interest on the dissemination of scientific publications on Twitter, the calculation of Twitter metrics is not free of challenges and limitations (Haustein, 2016; 2019). *Heterogeneity*, which refers to the diversity of acts and online events (Haustein, 2016), is one of the biggest challenges for altmetrics. Heterogeneity is not only observable across altmetric data sources in general, but appears in the reception of scientific publications on Twitter in particular. For example, there are various actions users can take to interact with scholarly contents on Twitter, such as originally tweeting, retweeting, replying, or liking tweets mentioning publications, among others (Haustein, Bowman, & Costas, 2016). There are multiple heterogeneous forms of co-occurrence that can happen in a single tweet, like hashtags, mentioned users, or URLs (Costas, de Rijcke, & Marres, 2017). Hence, when a Twitter mention is accrued, it is not just a simple number, but entails a multitude of information that refers to the different forms of interaction and exchange of information on Twitter. This lack of internal homogeneity (Wouters, Zahedi, & Costas, 2019) of Twitter metrics represents both a challenge as well as an opportunity, as it makes possible the further exploration of underlying patterns and user motivations (Sud & Thelwall, 2014) in their Twitter interactions with scientific publications.

Therefore, researchers are increasingly paying more attention to the content analysis of Twitter mentions and the behavioral analysis of Twitter users, going beyond the mere counting of tweets linking to scientific publications (Bornmann, 2014; Haustein, 2019). Twitter users' identities, motivations, and related interactions or engagement behaviors have been widely analyzed in order to improve the understanding of Twitter metrics in much more fine-grained manners (Holmberg et al., 2014; Haustein, Bowman, & Costas, 2015; Díaz-Faes, Bowman, & Costas, 2019). Nevertheless, rethinking the tweeting patterns and Twitter user behaviors in more detail comes with worries and problems that have aroused the concerns of researchers. By scrutinizing the patterns of tweeting of the top 10 most tweeted scientific dental papers, Robinson-Garcia et al. (2017) observed the mechanical nature of the bulk of tweeting behavior. This indicated that Twitter metrics based on simple counting of tweets run the risk of conflating multiple issues related with the tweeting activity, like obsessive single-user tweeting, duplicate tweeting, bots, and even human tweeting but devoid of original thought or engagement of the user with the paper in the tweet (Robinson-Garcia et al., 2017). Related concerns about Twitter data quality can be found in other studies as well (Thelwall et al., 2013b; Haustein et al., 2016).

*Consistency of altmetric data*

Data consistency is of great concern in studies of altmetric data. As Wouters, Zahedi, and Costas (2019) pointed out, among the characteristics of altmetric data, transparency and consistency are particularly essential for new indicators to be used for research evaluation. The lack of consistency is seen as one of the most noteworthy data quality challenges that all altmetric indicators have to confront (Haustein, 2016). Related research questions have been discussed from both conceptual and empirical perspectives, since article-level metrics emerged and were offered by several data providers with different data collection and integration principles (Chamberlain, 2013; Sutton, 2014; Zahedi, Fenner, & Costas, 2015).

Considering the strong dependency of altmetric data on commercial data providers, previous studies mainly focused on the consistency of various altmetric data among different data aggregators. For example, Ortega (2018) analyzed the coverage differences among Altmetric.com, PlumX and Crossref Event Data. These three altmetric data providers performed differently in each metric due to technical errors and extracting criteria, therefore, strategies of using specific services for particular metrics and combining different services for overall analysis were recommended. Meschede and Siebenlist (2018) also made a comparison between Altmetric.com and PlumX. They found that these two data aggregators achieved a moderate correlation overall but showed relatively weak consistency in some metrics, like Google+, Facebook, and News mentions. Zahedi and Costas (2018) presented an exhaustive study on the differences of data collection and reporting approaches among four major altmetric data providers, including Altmetric.com, PlumX, Lagotto, and Crossref Event Data. Similar results were confirmed and further explored in their study. More specifically, values of each metric provided by the different data aggregators obviously differed from each other because of their specific choices for the data collection and aggregation approaches. In a case study on the altmetric performance of papers published in the *Journal of the Association for Information Science and Technology* (*JASIST*) reported by Altmetric.com, PlumX, and Mendeley, the inconsistencies of metrics across data providers were observed by Bar-Ilan, Halevi, and Milojević (2019) in the same manner. Taken together, these results show that the data inconsistency at the data aggregator level is an important concern within the altmetric research community.

Moreover, as explained by Chamberlain (2013), altmetric data can be collected at different times, which potentially can also end up in obtaining different values of social media metrics, even when collected from the same source and for the same set of publications. This is one of the explanations for the differences in the data collected by different aggregators (Zahedi & Costas, 2018).

In this paper we introduce a different form of altmetric data inconsistency, related to the ever-changing nature of social media data, in which data records and social media events can easily be deleted by their creators, or users may abandon the social media platforms removing all their records from the platform. This form of inconsistency is therefore more related to the *stability* of altmetric data, and more specifically of Twitter metrics of publications. To the best of our knowledge, research on this type of inconsistency of Twitter metric data, as well as on their underlying causes, is still lacking in the social media metrics literature. In this paper we intend to fill this gap through a large-scale study of Twitter counts of publications collected at different times, focusing also on conceptualizing the potential reasons and risks that the observed instability may pose for the consistent calculation of Twitter metrics.

*Objectives*

The main objectives of this study are: (1) to investigate the stability of Twitter metrics by identifying Twitter mentions that have become unavailable over time, and (2) to explore the potential influence that these unavailable tweets may have on the overall Twitter metrics of publications. We addressed the following specific research questions:

Q1. What is the number and share of Twitter mentions of highly tweeted scientific publications in Altmetric.com that have become unavailable over time?

Q2. What are the most common reasons for tweets becoming unavailable?

Q3. To what extent do unavailable Twitter mentions influence the temporal stability of Twitter metrics of scientific publications?

Q4. Based on publications' unique Twitter dissemination structures consisting of original tweets, retweets, and retweeting links, is it possible to determine which scientific publications are at a higher risk of substantially decreased Twitter metrics when tweets become unavailable?

**Data and methods**

*Distribution of Twitter mention data recorded by Altmetric.com*

The Twitter mention data of scientific publications used in this study were extracted from the historical data files provided by Altmetric.com in 2017. Until October 2017, Altmetric.com has tracked and recorded nearly 43 million Twitter mentions for around 5.4 million unique scientific publications (namely Altmetric IDs). Altmetric.com provides two main indicators for measuring Twitter impact of scientific publications. One is the total number of tweets to the paper (TWS), the other is the number of unique Twitter users who have mentioned the paper (NUTU). Here we employ NUTU to present the distribution of Twitter mention data. Figures 1(a) and 1(b) show the NUTU distribution of all scientific publications recorded by Altmetric.com under a log-log scale and its probability density function (PDF), respectively. Distributions of several kinds of bibliometric data, such as citations (Brzezinski, 2015) and usage counts (Wang, Fang, & Sun, 2016), have been found to follow typical power law distributions, which is also observed in Figure 1(b) for Twitter mention data. Figure 1(b) is visualized based on the Python powerlaw package (Alstott, Bullmore, & Plenz, 2014), the distribution of unique Twitter users fits a power law distribution with $\alpha = 2.87$. Only a few scientific publications have attracted a large number of unique Twitter users, while the Twitter counts of most scientific publications are relatively low. In order to examine the stability of Twitter metrics, 1,154 scientific publications with at least 1,000 unique Twitter users (NUTU $\geq$ 1,000) were selected as our research objects. Until October 2017, these were the most tweeted scientific publications from the perspective of unique Twitter users involved, showing the highest impact on Twitter.

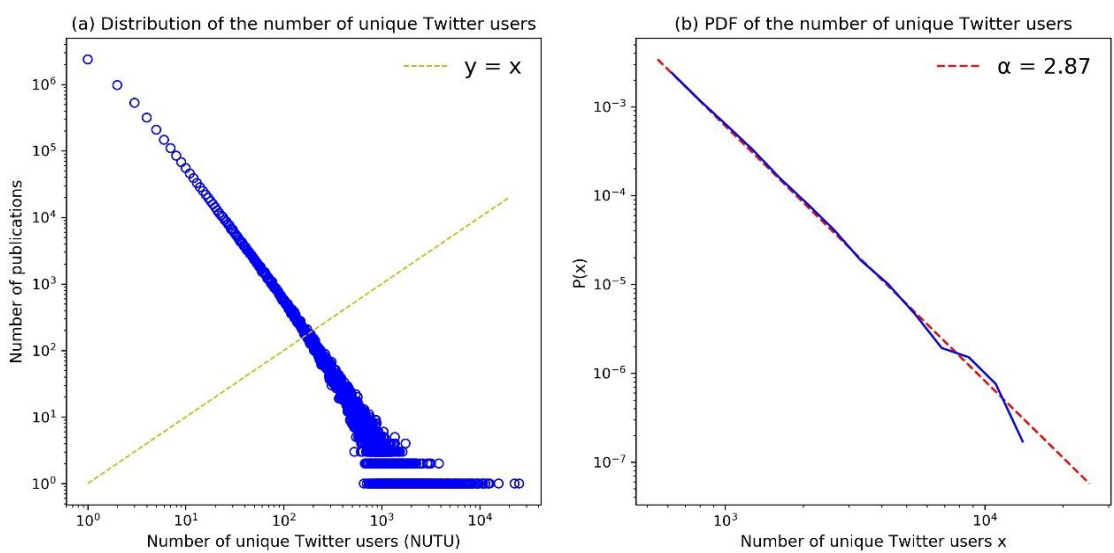

FIG. 1. Distribution and PDF of the number of unique Twitter users (NUTU).

*Availability of Twitter mentions of the most tweeted scientific publications*

Since 2016, Altmetric.com has made annual snapshots of its database available for researchers to study. These snapshots serve as an important reference point to study tweets that became unavailable at a later point in time. The snapshot data still provide evidence that a paper was tweeted even in the case when the tweet has been removed from Twitter, although the content and details of the tweet are not available any longer. For the 1,154 scientific publications with Twitter mentions posted by at least 1,000 unique Twitter users, all the tweet IDs (unique identifier of tweets) were collected from the data files provided by Altmetric.com (version: October 2017). In total, there are 2,643,531 unique tweet IDs related to the selected publications.

On the basis of the tweet IDs previously identified by Altmetric.com, in April 2019 we rechecked all the tweets through the Twitter API in order to examine of which tweets the status changed. For all tweets that are still available, detailed meta data can be acquired, and for those tweets that are no longer available, the Twitter API responds with respective error codes and error messages. Both unavailable tweet IDs and their error codes were recorded for further analysis. For the 2,643,531 Twitter mentions recorded by Altmetric.com until October 2017, a total of 378,766 (14.3%) were unavailable by April 2019.

*Indicators for describing Twitter dissemination structure*

In order to provide some understanding of the influence that unavailable tweets can have for the calculation of Twitter metrics, we study the *Twitter dissemination structures* of scientific publications. Twitter dissemination structure refers to the dissemination form of research outputs on Twitter over time, which is composed of *original tweets*, *retweets*, and the *retweeting links*. Original tweets are defined as Twitter mentions of scientific publications originally posted by Twitter users; retweets refer to the re-dissemination of original tweets by Twitter users; finally, the term retweeting links refers to the relationship between a specific original tweet and its following retweets, which is established when an original tweet is retweeted. The Twitter dissemination structure reveals how many original tweets a paper has accrued, how many retweets each original tweet has received, and how these original tweets and retweets connect over time.

As discussed before, a common Twitter metric for a scientific publication is the total count of tweets it has accumulated. However, the dissemination process of a scientific publication on Twitter is too intricate to be explained with a simple number. Studying the Twitter dissemination structures of scientific publications on Twitter can be seen as a more advanced approach to characterize the Twitter diffusion of scientific publications. *Originality* and *Concentration* are proposed as two dimensions for describing Twitter dissemination structures, which are based on the varieties that can be observed with scientific publications' original tweets, retweets, and their connections (i.e., retweeting links). Figure 2 illustrates four hypothetical examples of original tweet and retweet combinations in order to explain the two main dimensions for describing Twitter dissemination structures of publications. Blue nodes and yellow nodes represent original tweets and their related retweets, respectively. The four publications in the example (publication A, B, C, and D) do all have the same total number of Twitter mentions (TWS= 10). From the perspective of total tweet counts they show the same impact on Twitter, but they perform differently through the lens of *Originality* and *Concentration*.

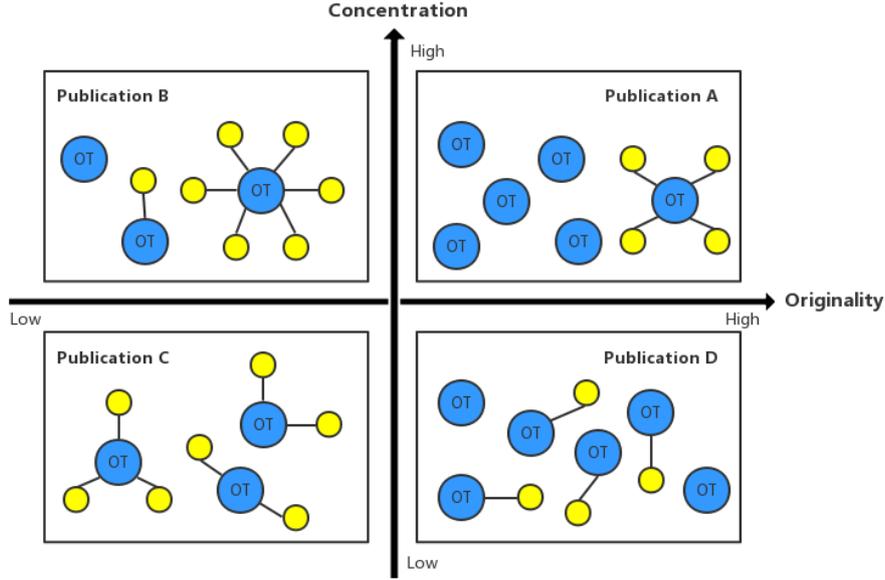

FIG. 2. Two dimensions for describing Twitter dissemination structures of publications, resulting in four different diffusion scenarios.

*Originality* is proposed to represent how many Twitter mentions of a specific scientific publication are posted originally by Twitter users rather than retweeting previous tweets. The more original tweets a publication has, the higher its degree of originality. The *Degree of Originality* (DO) of publication *x* is calculated as follows:

$$Degree\ of\ Originality_x = \frac{N(OT_x)}{TN(OT_x + RT_x)}$$

$N(OT_x)$ denotes the number of original tweets that publication *x* has received, while $TN(OT_x + RT_x)$ refers to the total number of Twitter mentions (including all original tweets and retweets) that publication *x* has accumulated. Essentially, DO reflects the proportion of original tweets a publication has received. In Figure 2, publication A (DO = 0.6) and publication D (DO = 0.6) fall into the category that has accumulated more original tweets; while publication B (DO = 0.3) and publication C (DO = 0.3) belong to the category that has received fewer original tweets.

*Concentration* is proposed to show the extent to which a publication's retweets are linked to its most retweeted original tweet. The more retweets concentrate on the most retweeted original tweet, the higher the publication's degree of concentration. The *Degree of Concentration* (DC) of publication *x* is given by:

$$Degree\ of\ Concentration_x = Max\left(\frac{N(RT_{OT_i})}{TN(RT_x)}\right)\ (i = 1,2,\dots,n)$$

$N(RT_{OT_i})$ denotes the number of retweets that original tweet $i$ $(i = 1,2,\dots,n)$ for publication *x* has received, $TN(RT_x)$ denotes the total number of retweets that publication *x* has accumulated. DC reflects the maximum percentage of retweets linking to (at least) a single original tweet. The higher the maximum percentage, the higher proportion of retweets concentrate on one single original tweet, while a low maximum percentage reflects a more disperse distribution of retweets. For publications without any retweet, their DCs are zero by default. For each publication in Figure 2, the proportions of retweets that every original tweet received are calculated and the maximum one is the DC of that publication. Therefore, the DCs of publication A and publication B are 1.0 and 0.86, respectively, with most retweets of these two publications concentrating on a certain original tweet; while for publication C (DC = 0.43) and publication D (DC = 0.25), retweets are distributed dispersedly.

All Twitter dissemination structures can be classified into the four categories in Figure 2 based on the two dimensions of *Originality* and *Concentration*. In order to study the dissemination structures of the highly tweeted publications selected for this study, their original tweets and retweets were distinguished at first. For Twitter mentions that are still available, the collected meta data indicate whether a tweet is an original tweet or a retweet, and in case it is a retweet, the tweet ID of its corresponding original tweet is returned as well, so that the retweeting links between original tweets and retweets can be identified. For Twitter mentions that are not available on Twitter any more, their status of original tweet and retweet, and their original tweet-retweet connections were established based on the data recorded by Altmetric.com, whenever this was possible. It should be noted that for some retweets, the corresponding original tweets are not always identified and recorded by Altmetric.com. Given that, in principle, the existence of a retweet relies on a corresponding original tweet, a possible explanation for the omission of original tweets is that during the data collection process by Altmetric.com, some retweets were identified and recorded first, and then the original tweets become unavailable before Altmetric.com could identify them, and therefore they were not included in the Altmetric.com data file. In those cases, we assumed that the original tweet must have existed at some point before the retweet. For the retweets without corresponding original tweets recorded, their original tweets are assumed for the sake of creating the retweeting links. Although these *assumed o*riginal tweets do not contribute to the total number of Twitter mentions of publications, they are considered to co-establish the Twitter dissemination structures of publications.

**Results**

The results section consists of three main parts: the first one explains the major reasons for the unavailability of Twitter mentions, and shows the distribution of unavailable Twitter mentions over the years. The second part presents the influence of unavailable Twitter mentions on Twitter metrics of scientific publications, and explores the possible causes for the highly unstable Twitter metrics through a case study. The last part focuses on the potential risks for publications with different Twitter dissemination structures of being unstable in Twitter metrics.

*Distribution of unavailable Twitter mentions*

Table 1 presents the number of unavailable Twitter mentions arranged by the specific *error codes* directly provided by the Twitter API. There are four main error codes that signal the unavailability of Twitter mentions. The major reason for the unavailability is that the tweet has been deleted, with around 54.7% of unavailable Twitter mention records falling into this category. The second major reason is that the Twitter user accounts have been suspended because of violation against Twitter rules[1], leading to the unavailability of all their tweets. This accounts for 25.9% of all errors returned, and is followed by the protection of tweets implemented by users[2]. Once a Twitter user has chosen this setting, unauthorized users cannot get access to their tweets (anymore), although the tweets themselves still exist. During our data collection, this error was found in the case of 16.7% of all unavailable Twitter mentions. Lastly, 2.7% of unavailable tweet IDs could not be found because the tweet IDs were directing to a page that does not exist anymore. It should be noted that in those cases where the tweet IDs are no longer existent (error codes 144 and 34), the related Twitter mentions about scientific publications are unrecoverable. Concerning unavailable tweet IDs due to user suspension or tweet protection (error codes 63 and 179), it is still possible that they become available to the public again once the suspended user accounts are unlocked or the users cancel the protection of their tweets. Nevertheless, whether such reversion will take place is uncertain, thus the unavailability of these tweet IDs still has a negative effect on the stability of the Twitter metrics.

---

[1] See more information about suspended Twitter accounts at: https://help.twitter.com/en/managing-your-account/suspended-twitter-accounts
[2] See more information about public and protected tweets at: https://help.twitter.com/en/safety-and-security/public-and-protected-tweets

TABLE 1. Numbers of unavailable Twitter mentions and reasons for their unavailability.

| Error code | Twitter Error message | Description | N | P |
| --- | --- | --- | --- | --- |
| 144 | No status found with that ID. | The requested Tweet ID is not found (if it existed, it was probably deleted). | 207,147 | 54.7% |
| 63 | User has been suspended. | The user account has been suspended and information cannot be retrieved. | 98,194 | 25.9% |
| 179 | Sorry, you are not authorized to see this status. | Thrown when a Tweet cannot be viewed by the authenticating user, usually due to the Tweet's author having protected their Tweets. | 63,393 | 16.7% |
| 34 | Sorry, that page does not exist. | The specified resource was not found. | 10,032 | 2.7% |
| Total | | | 378,766 | 100.0% |

Altmetric.com started tracking Twitter data from October 2011 onwards [3]. Figure 3 shows the distribution of the Twitter mentions of the 1,154 most tweeted scientific publications over the years, as well as of the unavailable Twitter mentions. Each bar in Figure 3 presents the total number of Twitter mentions with posting date information every year, and the percentage of unavailable Twitter mentions is represented by the lined segments in the bars, and numerically listed in brackets. Older Twitter mentions (e.g., from years 2011, 2012, or 2013) exhibit higher proportions of unavailable tweets, suggesting that the longer the time between the tweet and the data collection, the larger the chances of finding unavailable tweets.

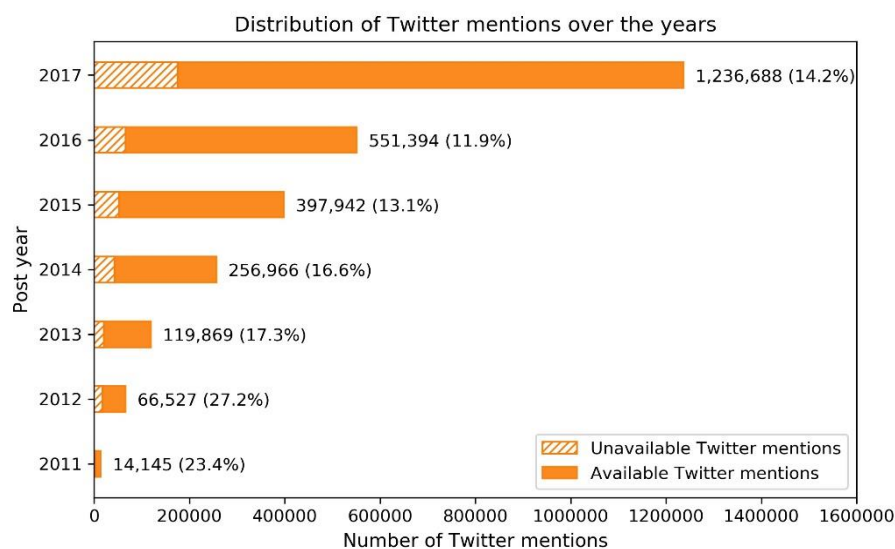

FIG. 3. Distribution of Twitter mentions over the years. (Share of unavailable tweets per year listed in brackets)

---

[3] When did Altmetric start tracking attention to each attention source? Retrieved from
https://help.altmetric.com/support/solutions/articles/6000136884-when-did-altmetric-start-tracking-attention-to-each-attention-source-

*Influence of unavailable Twitter mentions on the stability of Twitter metrics*

Figure 4 shows the total number of Twitter mentions (blue line) and still available Twitter mentions (orange line) for 1,154 Altmetric IDs. The *Twitter unavailability rate*, namely the percentage of unavailable Twitter mentions of each scientific publication, is presented as a yellow dashed line. For clearer visualization, the 1,154 publications are divided into three parts in the order of their total number of Twitter mentions and shown in Figure 4(a), 4(b), and 4(c), respectively. All highly tweeted publications have a certain number of unavailable tweets, and the amounts vary greatly across publications. Peaks of yellow dashed line represent those publications with a large share of unavailable Twitter mentions. Due to these high unavailability rates, it can be argued that the Twitter metrics of the corresponding publications are unstable.

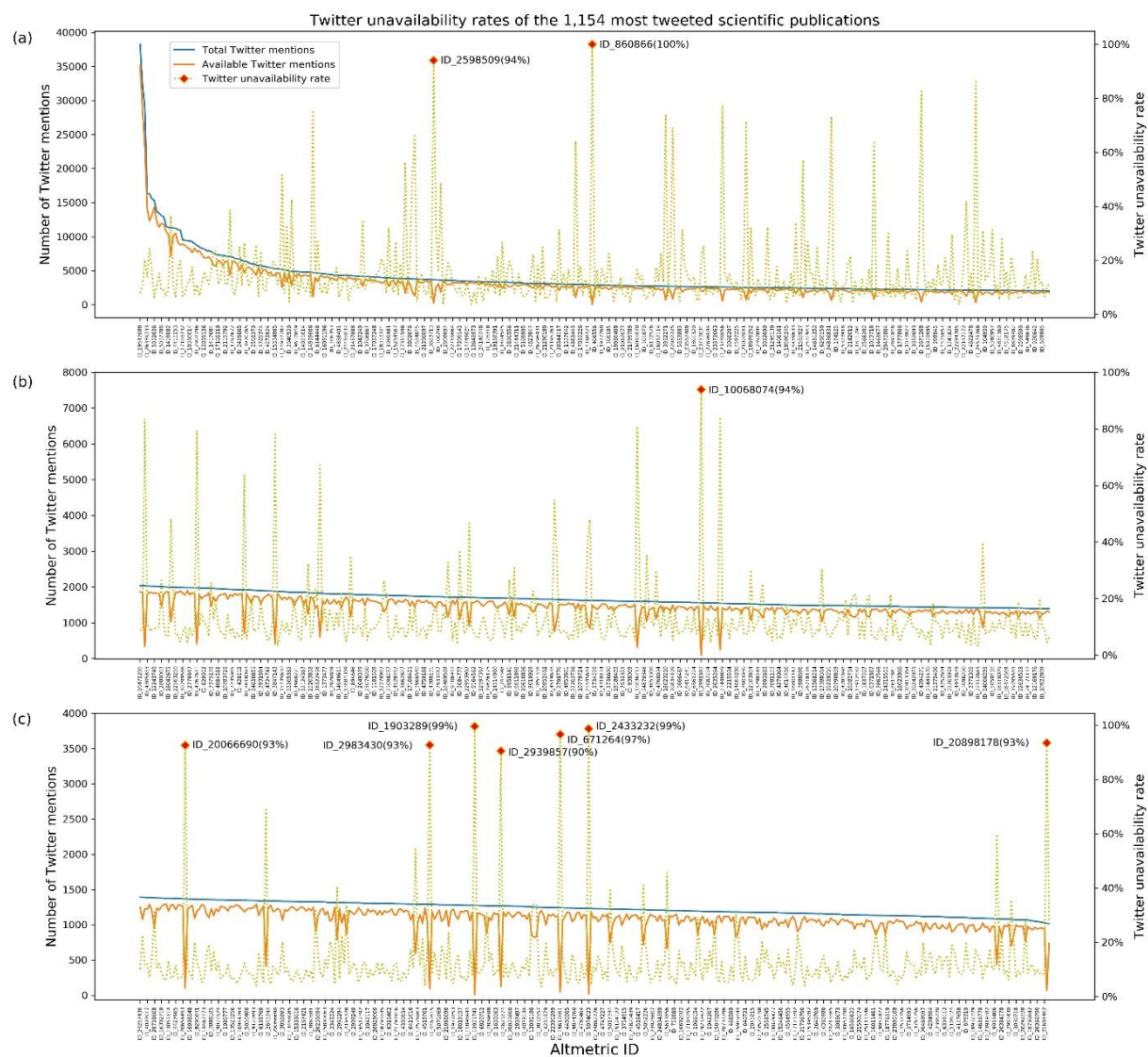

FIG. 4. Twitter unavailability rates of the 1,154 most tweeted scientific publications.

In order to investigate potential causes for the high Twitter unavailability rates of some publications, we selected the top 10 Altmetric IDs with the highest unavailability rate of Twitter mentions as a case study. In Figure 4, these top 10 Altmetric IDs are highlighted with red diamonds. The Twitter metrics of these scientific publications are most seriously affected by unavailable Twitter mentions, since their Twitter metrics sharply decreases, causing the *demotion* of these publications as highly tweeted publications. Table 2 presents details of their unavailable Twitter mentions from the perspectives of original tweets and retweets in detail. Total number of tweets to the paper, number of recorded original tweets, number of unavailable tweets, Twitter unavailability rate, number of unavailable original tweets, number of unavailable retweets, and maximum number of unavailable retweets related to an original tweet are calculated to reflect the composition of unavailable Twitter mentions.

TABLE 2. Top 10 Altmetric IDs with the highest unavailability rate of Twitter mentions.

| Altmetric ID | DOI | TWS | N_OT | N_UnT | TUnR | N_UnOT | N_UnRT | Max(N_UnRT) |
|---|---|---|---|---|---|---|---|---|
| 860866 | 10.1088/1475-7516/2008/10/036 | 2891 | 1 | 2891 | 100.0% | 1 | 2890 | 2890 |
| 1903289 | 10.2337/diacare.27.2007.s111 | 1274 | 3 | 1268 | 99.5% | 0 | 1268 | 1268 |
| 2433232 | 10.1056/nejmoa1315231 | 1230 | 11 | 1213 | 98.6% | 0 | 1213 | 1213 |
| 671264 | 10.1056/nejmoa1109017 | 1241 | 23 | 1198 | 96.5% | 0 | 1198 | 1190 |
| 2598509 | 10.1080/17439884.2014.942666 | 3659 | 122 | 3440 | 94.0% | 4 | 3436 | 3319 |
| 10068074 | - | 1563 | 94 | 1467 | 93.9% | 17 | 1450 | 1426 |
| 20898178 | 10.1097/adm.0000000000000324 | 1017 | 34 | 950 | 93.4% | 0 | 950 | 950 |
| 2983430 | 10.2139/ssrn.2536258 | 1290 | 76 | 1195 | 92.6% | 41 | 1154 | 151 |
| 20066690 | 10.1038/nrmicro.2017.40 | 1367 | 10 | 1265 | 92.5% | 1 | 1264 | 1253 |
| 2939857 | - | 1266 | 86 | 1145 | 90.4% | 43 | 1102 | 248 |

Note: TWS = total number of tweets to the publication; N_OT = number of recorded original tweets; N_UnT = number of unavailable Twitter mentions; TUnR = Twitter unavailability rate; N_UnOT = number of unavailable original tweets; N_UnRT = number of unavailable retweets; Max(N_UnRT) = maximum number of unavailable retweets related to a single original tweet. Altmetric IDs 10068074 and 2939857 are publications without DOI registered.

More than 90% of the Twitter mentions of these 10 scientific publications are unavailable, and almost all unavailable Twitter mentions are retweets. Therefore, for unavailable retweets of each publication, we explored the reasons for the high unavailability rate by calculating the maximum number of unavailable retweets related to a single original tweet for each publication. The results indicate that except for two Altmetric IDs (2983430 and 2939857), most unavailable retweets concentrate on a specific original tweet. For example, Altmetric ID 860866 has 2,891 Twitter mentions in total, consisting of only one original tweet and 2,890 retweets related to that original tweet. Therefore, when the original tweet became unavailable, according to the rules of Twitter[4], all its related retweets that used Twitter's native "retweet" functionality turned unavailable as well, virtually decreasing the Twitter metrics of the publication to zero. The same happens to other Altmetric IDs with most unavailable retweets concentrating around an original tweet that became unavailable. In Table 2, there are four Altmetric IDs where the number of unavailable original tweets equals zero. In fact, the unavailable retweets of these four publications direct to an unavailable original tweet as well according to our manual check. The zero values of N_UnOT are caused by the omission of original tweets in Altmetric.com's data files as we mentioned before. Based on these results we can state that the unavailability of an original tweet leads to the unavailability of a large number of retweets concentrating on it. This is the main reason for the high Twitter unavailability rates of publications listed in Table 2.

*Twitter unavailability rates of publications with different Twitter dissemination structures*

In order to further investigate the potential influence of different Twitter dissemination structures on the (in)stability of Twitter metrics, we calculated the *Degree of Originality* (DO) and *Degree of Concentration* (DC) for 1,154 sample scientific publications, with the distribution shown in Figure 5. Each dot represents a publication, and its color is determined by the Twitter unavailability rate shown in the color bar on the right. The dashed vertical and horizontal lines indicate the median DO (0.284) and median DC (0.203) of all publications, respectively. Moreover, the top 10 publications with the highest unavailability rate of Twitter mentions listed in Table 2 are marked by stars to highlight their location in the scatter plot. Most publications with high Twitter unavailability rates are located at the upper left part, especially for the eight starred publications with the highest unavailability rates. Their Twitter dissemination structures have very low DO and quite high DC, which means that once an original tweet with lots of retweets linking to it has been removed, most of that publication's Twitter mentions become unavailable. This results in the collapse of its Twitter metrics. Some publications at the left lower part, namely those with both low DO and low DC, also show quite a high unavailability rate of Twitter mentions. This kind of publications only have few original tweets but most of them received some retweets. Here, the distribution of retweets is more balanced, meaning that the risk of losing most of the retweets received once the original tweet becomes unavailable is not as high as for the publications at the upper left part. However, if the few original tweets received come from a specific Twitter user, and that user account is suspended, or that user decides to protect their tweets, the stability of Twitter metrics of those publications would be seriously affected as well. This is the case with the two starred publications at the left lower part. There are less publications with high Twitter unavailability rates in the right part. Publications in this part accumulated more original tweets, so they have less retweets that rely on the existence of original tweets. Throughout all four fields, the Twitter metrics of publications with high DO and low DC (right lower part) seem to be the most stable, since their dissemination structures consist of more independent original tweets and more decentralized retweets, which lowers the risk of losing a lot of Twitter records caused by the unavailability of several highly retweeted original tweets.

---

[4] See more information about rules of tweet deletion at: https://help.twitter.com/en/using-twitter/delete-tweets

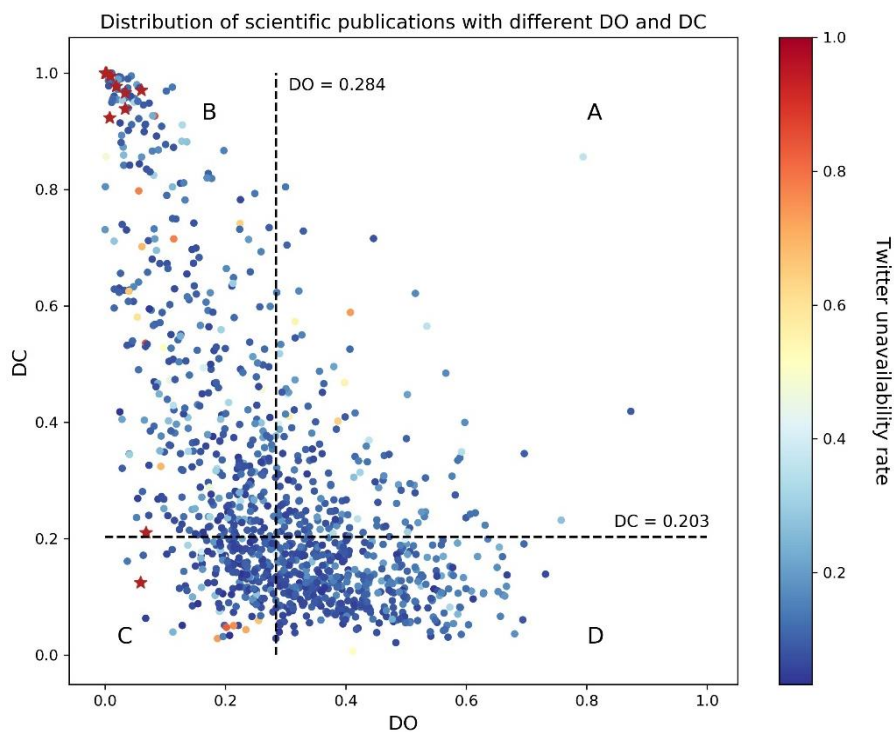

FIG. 5. Distribution of 1,154 scientific publications with different DO and DC.

The dashed lines in Figure 5 represent the median value of DO and DC respectively and classify the publications included into four groups (A, B, C, and D). This is in correspondence with the four categories we introduced in Figure 2. The distribution of Twitter unavailability rates of these four groups of publications is shown in an associative plot (Figure 6, box plot and violin plot). With all four groups, most Twitter unavailability rates locate below 0.2, suggesting that most publications in these four groups have less than 20% of their Twitter mentions unavailable, their Twitter metrics are relatively stable regarding minor losses. However, the amount and distribution of outliers vary across groups. Group B and Group C have more outliers that hold extremely high Twitter unavailability rates, while those of Group A and Group D are fewer. Besides, most outliers of the latter are below 0.6; by contrast, Groups B and C have lots of outliers higher than 0.8, especially Group B. These results are in line with what we observed in Figure 5. Although most publications with different Twitter dissemination structures keep a relatively low Twitter unavailability rate, publications with extremely unstable Twitter metrics are more likely to occur when they have fewer original tweets and more concentrated retweets (Group B) or less original tweets and relatively deconcentrated retweets (Group C).

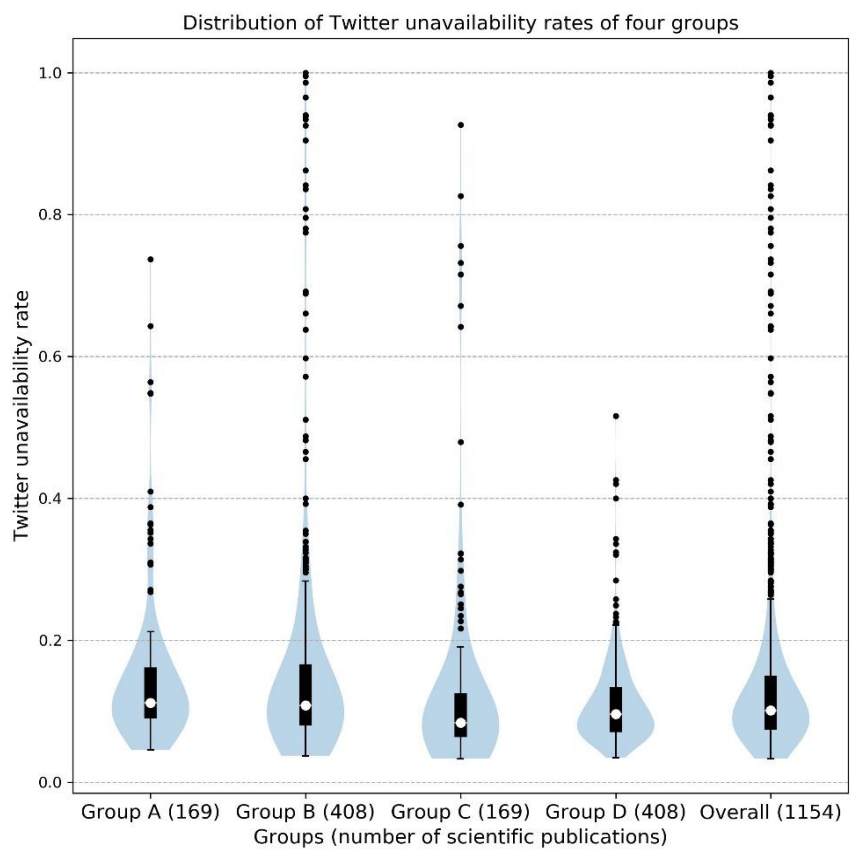

FIG. 6. Distribution of Twitter unavailability rates of the four groups with different DO/DC characteristics.

## Discussion

*The possible instability of Twitter metrics*

Data consistency is essential for the measurement of impact in a sustainable and stable manner. In the context of altmetrics, data consistency is significantly affected by the dynamic nature of events (Haustein, 2016). Conceptually speaking, citations, once given, cannot disappear. Therefore, the decrease of citation counts of a specific publication is very rare, and mostly caused by technical issues (e.g., changes in the coverage of the database, changes in the citation matching algorithms, etc.). For this reason, citation-based metrics of scholarly outputs are relatively stable over the course of time. On the other hand, there are no barriers for Twitter users to post a tweet or retweet, neither to delete a tweet or to cancel a retweet. A previously existing Twitter mention might become unavailable to the public for various reasons, and can no longer be identified or reused by following data aggregators and users, leading to the instability of Twitter counts of mentions to scientific publications. The same situation also happens to other altmetric indicators, for instance, Mendeley readership (Bar-Ilan, 2014). The number of Mendeley readership could decrease when users remove older references from their libraries (Zahedi, Costas, & Wouters, 2017), leading to the instability of readership counts as time goes by. Moreover, in the study of availability of blogs and news links, Ortega (2019) observed that a considerable share of links in Altmetric.com and PlumX are broken due to the disappearance of some third parties that supply news and blog events, thereby making those news and blog records unavailable and which therefore cannot be audited.

In this study we checked the availability statuses of over 2.6 million Twitter mentions of the 1,154 most tweeted scientific publications recorded by Altmetric.com up to October 2017, to examine their Twitter unavailability rates, that is, the extent of Twitter mentions having become unavailable to the public. The status and reasons for unavailability were retrieved in April 2019. Our results indicate that for these most tweeted publications, around 14.3% of their Twitter mention records have become unavailable to the public. Twitter mentions that have been posted for a long time show a higher proportion of unavailability. Thus, the potential risk of Twitter mentions being unavailable for different reasons increases over time. Nevertheless, because Twitter users have become more active in sharing scientific information in recent years, the absolute number of unavailable Twitter mentions in 2017 is much higher than before. User deletion is the main reason for this high unavailability rate, accounting for 54.7% of unavailable Twitter records, followed by suspension and protection of Twitter user accounts (accounting for 25.9% and 15.7%, respectively).

Twitter unavailability rates vary markedly among scientific publications, hence influencing their Twitter metrics to different extents. In our study, all selected highly tweeted publications have a certain share of Twitter mentions unavailable at the time of data collection, and most of them have less than 20% of Twitter mentions that have become unavailable to the public. However, there are many publications that show extremely high unavailability rates. In our case study of the top 10 publications with the highest Twitter unavailability rates, over 90% of Twitter mentions directing to them have become unavailable. For these scientific publications, their Twitter metrics are among the highest when they were recorded by Altmetric.com, but if the unavailable Twitter mentions would be excluded from the counts, the overall Twitter counts of these publications would plummet dramatically. This is even more concerning given that Twitter data show a fast accumulation speed. In general, over 80% of Twitter data are accumulated within the first year after publication (Fang & Costas, 2018). This means that once the Twitter metrics of a relatively old publication has been affected by unavailable tweets, it is difficult for the publication to receive as many Twitter mentions as it had before to recover its Twitter metrics again. In this case, for publications that have been published for a long time, in general the loss of Twitter mentions is irreversible. What is more important is that those unavailable Twitter mentions cannot be detected and counted by other data aggregators that never recorded them before, which might exacerbate the inconsistency among Twitter data recorded by different data aggregators.

*The influence of different forms of Twitter dissemination structures*

In order to further explore the underlying reasons for high Twitter unavailability rates, we analyzed the Twitter dissemination structures of scientific publications based on the composition of their original tweets, retweets, and the connections between them. *Originality* and *Concentration* were introduced as two dimensions to classify these Twitter dissemination structures. Furthermore, *Degree of Originality* (DO) and *Degree of Concentration* (DC) were proposed as two new measures to describe how many original tweets a publication has received (DO) and to what extent retweets concentrate around these original tweets (DC). On the basis of these two indicators, we found that scientific publications showing a relatively low *Degree of Originality* and a relatively high *Degree of Concentration* are at a greater risk of losing larger numbers of Twitter mentions. This is because once a highly retweeted original tweet becomes unavailable, all its related retweets also become unavailable, generating a dramatic decrease in the overall Twitter metrics of the publication in question. In addition, some publications with extremely unstable Twitter metrics also show relatively low *Degree of Originality* and relatively low *Degree of Concentration*. In most cases, this is because the few original tweets were posted by the same user account, namely those user accounts who tweeted the same paper repeatedly, as observed by Robinson-Garcia et al. (2017). If the Twitter user sending original tweets repeated times is suspended, all of his original tweets become unavailable, and so do the related retweets. By comparison, among publications with relatively high *Degree of Originality* there are a few showing extremely unstable Twitter metrics, particularly when the *Degree of Concentration* is low. The high *Degree of Originality* lowers to some extent the risk of losing the bulk of Twitter records.

Given the diversity of users and complexity of engagement behaviors that happen on Twitter, the dissemination processes of scientific publications on Twitter are sophisticated, Twitter metrics can help to unveil such diversity and complexity (Haustein, 2019). The study of Twitter dissemination structures does not only contribute to the identification of publications that may suffer from a stronger vulnerability of losing their Twitter counts, but also sheds light on the possibilities of measuring performance of scientific publications on Twitter in a more fine-grained manner. The total count of Twitter mentions is one of the most common Twitter measures, but as we presented in Figure 2, publications with the same total Twitter mention counts might perform differently from the point of view of their Twitter dissemination structures. Didegah, Mejlgaard, and Sørensen (2018) studied the number of original tweets and retweets of publications and their qualities across different subject fields. But beyond this kind of statistics, it is relevant to organize these data to reveal the overall picture of Twitter dissemination structures of research outputs. Twitter impact is not only about how many times a paper has been tweeted, but also about how it was tweeted. The reconstruction of the Twitter dissemination structure provides a partial answer to this question. Based on the Twitter dissemination structure, it is possible to unravel the underlying dissemination patterns and networks of publications that hide behind the total statistical numbers, with the latter compounding different types of Twitter mentions and their relationships in a simple way. As a result, the Twitter dissemination structure is supposed to contribute to a better understanding of the performance of publications on Twitter.

In future research, we will further optimize the indicators for describing Twitter dissemination structures. For example, in this paper the *Degree of Concentration* is calculated based on the maximum percentage of retweets concentrating on a single original tweet. This method derived from the case study of the top 10 publications with the highest Twitter unavailability rates, it is also a way with the advantage of simplicity. We will introduce multiple calculation methods for measuring the *Degree of Concentration* at both tweet and Twitter user levels in future studies. Particularly at the Twitter user level, in addition to taking the retweeting relationships among users into account, the status, the degree of activity, and diverse Twitter user profiles are expected to be considered to establish more fine-grained Twitter dissemination structures. Moreover, we will explore possible applications of Twitter dissemination structures in the measurement of the Twitter performance and impact of scientific outputs.

*Overall situation of the stability of Twitter metrics*

Besides rechecking the Twitter mentions of the 1,154 most tweeted publications presented above, in September 2019 we rechecked the statuses of all Twitter mentions recorded by Altmetric.com in the historical data files (version: October 2017) to reveal the overall situation of the stability of Twitter metrics for nearly 5.4 million publications. Results show that among the over 42.5 million unique recorded Twitter mentions, about 13.0% of them have become unavailable. Accordingly, the overall Twitter unavailability rate is slightly lower than that of the sample of highly tweeted publications (14.3%).

For understanding the overall influence of unavailable tweets on Twitter metrics at the publication level, Spearman correlation analyses between the total number of recorded Twitter mentions and the number of available Twitter mentions during data rechecking are conducted for both the sample of the 1,154 most tweeted publications and all recorded publications in Altmetric.com. For both datasets, these two numbers are highly correlated ($r_s = 0.91$ for the most tweeted publications, and $r_s = 0.93$ for all publications), which means that the majority of publications kept a relatively stable Twitter metrics over time. This result is in line with the distribution of Twitter unavailability rates we observed for the most tweeted publications, with most publications having less than 20% of tweets unavailable and a limited share of publications showing extremely unstable Twitter metrics.

It should be noted that although the value of Altmetric.com database snapshots is obvious for studying changes in altmetrics over time, due to the Twitter restrictions, Altmetric.com is no longer providing tweets that have been removed from Twitter, and researchers are now required to delete all unavailable tweets from their locally hosted snapshot files[5]. This implies that unavailable tweets cannot be studied in related future research. Moreover, except for tweet IDs and Twitter user IDs, Altmetric.com will no longer provide the content of Twitter mentions of publications in its snapshots, ensuring that the detailed information of potential unavailable tweets not be kept in the historical data files.

*Limitations*

There are some limitations that should be acknowledged in this study. First, as we mentioned in the data and methods section, there exist some retweets without corresponding original tweets recorded by Altmetric.com. Given that the existence of an original tweet is the basis of its following retweets, we assumed that there are some original tweets to complete the retweeting relationship, meaning that we had to work with "assumed" data instead of actual data. Secondly, for deleted original tweets, it would be interesting to analyze the motivations of users. However, this question is not further discussed in our paper because of the lack of traceable evidence and the Twitter restrictions on deleted content. Lastly, Twitter dissemination structures were analyzed only from the perspective of the connections of different types of tweets (original tweets and retweets), whereas diversity of Twitter users in the Twitter dissemination process might be another factor that has an influence on the stability of Twitter metrics. In the case of some publications with relatively low *Degree of Originality* and low *Degree of Concentration*, we could show that the reason why they have extremely high unavailability rates is that the few original tweets were posted by the same user account. Therefore, the composition of Twitter users involved and their identities should be further explored in the future, especially in the light of bot accounts playing a major role in the science communication landscape on Twitter (Didegah, Mejlgaard, & Sørensen, 2018).

**Conclusions**

This study examines the stability of Twitter metrics of scientific publications by rechecking the statuses of their Twitter mentions. For over 2.6 million Twitter records of the 1,154 most tweeted publications recorded by Altmetric.com until October 2017, about 14.3% of them have become unavailable by April 2019. The main reason for the high unavailability rate is deletion of tweets, followed by suspension and protection of Twitter user accounts. The stability of Twitter metrics varies among publications, most of them have Twitter unavailability rates of less than 20%, but there are some publications showing extremely high unavailability rates. The potential influence of Twitter dissemination structures on the stability of Twitter metrics is investigated. *Degree of Originality* and *Degree of Concentration* are proposed to describe Twitter dissemination structures based on original tweets, retweets, and original tweet-retweet connections. Twitter metrics of publications with relatively low *Degree of Originality* and relatively high *Degree of Concentration* are at greater risk of becoming highly unstable. Building on that, we discuss the stability and persistency of Twitter metrics of scientific publications and the potential risks they can be subject to. Thus, our study underlines the importance of distinguishing dissemination structures in the context of Twitter-based indicators.

**Acknowledgements**

Zhichao Fang is financially supported by the China Scholarship Council (Grant No. 201706060201). Rodrigo Costas is partially funded by the South African DST-NRF Centre of Excellence in Scientometrics and Science, Technology and Innovation Policy (SciSTIP). The authors thank Prof. Paul Wouters (Leiden University) for valuable suggestions, thank Stacy Konkiel from Altmetric.com for her explanation of the snapshot files and their solutions to historical Twitter records, and also thank the three anonymous reviewers for their helpful suggestions.

---

[5] Extracted from personal communication with Stacy Konkiel from Altmetric.com.